\documentclass[aps,pra,twocolumn,amsmath,amssymb,nofootinbib,superscriptaddress]{revtex4}
\usepackage[english]{babel}
\usepackage{latexsym}
\usepackage{graphics}
\usepackage{graphicx}
\usepackage{epsfig}
\usepackage{bm}
\usepackage{amsmath}
\usepackage{amssymb}
\usepackage{amsthm}
\usepackage{dcolumn}
\usepackage{subfig}
\usepackage{float}
\usepackage{tikz} 
\usetikzlibrary{calc}
\usepackage{pgfplots}
\pgfplotsset{compat=newest, width=8 cm}
\usepgfplotslibrary{groupplots}
\usetikzlibrary{arrows.meta}
\usepgfplotslibrary{fillbetween}
\usepgfplotslibrary{polar}
\usepackage{pgfplotstable}
\usepackage{hyperref}
\graphicspath{{Images/}}
\usepackage{comment}
\usepackage{epstopdf}
\usepackage{cancel}
\usepackage{cleveref}
\usepackage{enumerate}
\usepackage{braket}
\usepackage{appendix}
\usepackage{standalone}
\usepackage{capt-of}
\newcommand{\bs}{\boldsymbol}
\newcommand{\ovl}{\overline}

\DeclareUnicodeCharacter{2009}{{ }}

\theoremstyle{remark}

\pgfplotsset{colormap={violet}{rgb255=(25,25,122) rgb255=(238,140,238) color=(white)}}

\makeatletter
\pgfmathdeclarefunction{erf}{1}{%
  \begingroup
    \pgfmathparse{#1 > 0 ? 1 : -1}%
    \edef\sign{\pgfmathresult}%
    \pgfmathparse{abs(#1)}%
    \edef\x{\pgfmathresult}%
    \pgfmathparse{1/(1+0.3275911*\x)}%
    \edef\t{\pgfmathresult}%
    \pgfmathparse{%
      1 - (((((1.061405429*\t -1.453152027)*\t) + 1.421413741)*\t 
      -0.284496736)*\t + 0.254829592)*\t*exp(-(\x*\x))}%
    \edef\y{\pgfmathresult}%
    \pgfmathparse{(\sign)*\y}%
    \pgfmath@smuggleone\pgfmathresult%
  \endgroup
}
\makeatother

\pgfplotsset{
        compat=1.12,
        /pgf/declare function={
            f(\x) = 1+sqrt(2*\x/pi)*exp(-0.5/(\x))/(1+erf(sqrt(0.5/(\x));
        },
        /pgf/declare function={
            density01(\x) = exp(-(\x-1)*(\x-1)/(2*0.1))/(sqrt(0.5*pi*0.1)*(1+erf(sqrt(0.5/0.1))));
        },
        /pgf/declare function={
            density20(\x) = exp(-(\x-1)*(\x-1)/(2*20))/(sqrt(0.5*pi*20)*(1+erf(sqrt(0.5/20))));
        },
        /pgf/declare function={
            density1(\x) = exp(-(\x-1)*(\x-1)/(2*1))/(sqrt(0.5*pi*1)*(1+erf(sqrt(0.5/1))));
        },
    }

\begin{document}

\title{Generating entangled polaritonic condensates by pumping with entangled pairs of photons}
\author{N.A. Asriyan}
\email{norair.asriian@uni.lu}
\affiliation{N.L. Dukhov Research Institute of Automatics (VNIIA), Moscow 127030, Russia}
\affiliation{Complex Systems and Statistical Mechanics, Department of Physics and Materials Science, University of Luxembourg, 30 Avenue des Hauts-Fourneaux, Esch-sur-Alzette L-4362, Luxembourg}

\author{A.A. Elistratov}
\affiliation{N.L. Dukhov Research Institute of Automatics (VNIIA), Moscow 127030, Russia}

\author{A.V. Kavokin}
\affiliation{Russian Quantum Center, Skolkovo, Moscow 143025, Russia}
\affiliation{Abrikosov Center for Theoretical Physics, Moscow Institute of Physics
and Technology, Institutskiy per. 9, Dolgoprudny, Moscow Region 141701, Russia}
\affiliation{School of Science, Westlake University, 18 Shilongshan Road, Hangzhou, Zhejiang Province 310024, China}

\begin{abstract}
    We investigate the steady state of two single-mode, uniform, spatially separated polaritonic condensates exposed to resonant pumping with entangled pairs of photons in the presence of incoherent exciton reservoirs created by nonresonant optical pumping. We demonstrate the principal possibility of driving the system to an entangled state despite its exposure to noises arising from the excitonic reservoirs and photon leakage through the microcavity mirrors. We estimate the minimum pair-induced polariton influx required to drive the system into a steady state that violates the partial-transpose criterion for entanglement. Furthermore, we trace the evolution of the system after a sudden switch-off of the entangled pumping. Our analysis provides estimates for the entanglement lifetime in a system of two exciton-polariton condensates.
\end{abstract}

\maketitle
\section{Introduction}
Entanglement is one of the most remarkable quantum phenomena at the microscopic level. However, its observation and manipulation at the macroscopic level remain challenging. Recent experiments with ultracold atomic gases have demonstrated that Bose--Einstein condensates are promising candidates for this purpose. Initially, it was demonstrated that distinct spatial modes of the same condensate can exhibit quantum correlations  ~\cite{doi:10.1126/science.aao1850}. More recently, an entangled state of two macroscopic condensates was achieved ~\cite{PhysRevX.13.021031}.

When it comes to excitonic polaritons, there is still much more to do to obtain an experimental demonstration of macroscopic entanglement. Recent theoretical work sheds light on the possible research directions. In particular, proposals have considered the entanglement between different polarization modes of the same condensate~\cite{PhysRevA.104.013318}, generation and detection of entanglement between spatially separated parts of a polariton condensate~\cite{PhysRevA.108.053301}, and, more recently, entanglement between coupled trapped polariton condensates driven by control pulses~\cite{Kavokin2026Classical}.

Entanglement has also been theoretically explored in related semiconductor microcavity systems. In particular, steady-state entanglement has been predicted between excitonic and exciton-polariton modes in exciton-optomechanical systems~\cite{Zuo2024}, as well as between two spatially separated exciton modes driven by squeezed optical fields~\cite{EsQars2025}.

In contrast, experimental progress remains more limited. Quantum correlations between a photon and a polariton have been demonstrated~\cite{cuevasFirstObservationQuantized2018}, but entanglement of two macroscopic polariton condensates has not yet been observed.

Despite the fundamental importance of achieving entanglement with exciton-polariton condensates, there may also be practical benefits for modern quantum technologies~\cite{kavokin_review_polariton_2022}. A key requirement for using this physical platform as a component of a qubit is the robustness of entanglement with respect to environmental noise. For superconducting, semiconductor-based, and trapped-ion qubits, low temperatures are essential for suppressing environmental noise. Given the low particle mass, and consequently, the relatively high temperature for persisting collective quantum behavior, exciton-polaritons might impose less strict temperature requirements for entanglement preservation. Several proposals have explored how polariton condensates can be used to realize qubits~\cite{ghosh_quantum_2020, PhysRevResearch.3.013099}.

In general, the challenge to demonstrate macroscopic entanglement of two clouds is significant because of the highly nonequilibrium nature of the polaritonic system, which elevates the level of thermal and quantum noise. In this paper, we analyze the possibility of the entanglement of polariton condensates from the quantum optical perspective. Aiming at a proof-of-principle demonstration, we resort to a brute-force method: directly injecting already entangled photons from an external source. Related schemes for transferring photonic entanglement to spatially separated matter systems have been put forward for atomic Bose--Einstein condensates~\cite{Gullo2010} and distant qubits~\cite{Agusti2022,Agusti2026}. Here, we ask whether such pumping with entangled photon pairs can similarly generate entanglement between two independent polariton condensates despite the intrinsic gain and loss fluctuations associated with their driven-dissipative nature. In other words, we consider parametrically coupled bosonic modes exposed to a noisy environment.

Our goal is to find conditions under which the condensates will be entangled and to estimate the lifetime of entanglement in the noisy environment of pump and decay reservoirs. In the next section, we introduce the model and discuss the mean-field behavior of the pumped condensates. Then, in Section \ref{sec:statstate}, follows an analysis of the stationary state of two condensates above the condensation threshold with fluctuations taken into account. The main results of the paper are reported in Section \ref{sec:test}, where we derive, in general form, sufficient conditions on the effective pair-drive strength that lead to an entangled state of the condensates. These results are followed by an estimate of the entanglement lifetime in Section \ref{sec:lifetime} and a discussion of the implications of the results in Section \ref{sec:disc}.

\section{Model}\label{sec:model}

\begin{figure}[htp]
    \centering
    \includegraphics{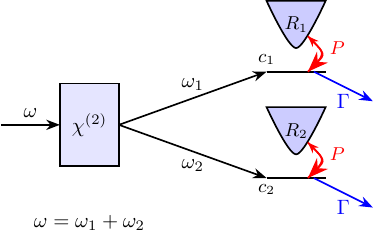}
    \caption{A possible schematic of the proposed setup. A nonlinear medium with quadratic susceptibility $\chi^{(2)}$ is pumped by a laser of frequency $\omega=\omega_1+\omega_2$. It generates entangled photonic pairs resonant with the two polariton condensate modes $c_1$ and $c_2$. The two condensates are spatially separated and coupled to independent incoherent exciton reservoirs $R_1$ and $R_2$, which provide nonresonant pumping with flux $P$. The condensate modes experience photon leakage through the cavity mirrors at rate $\Gamma$.}
    \label{fig:illustration}
\end{figure}

The schematic of the model we use is shown in Fig. \ref{fig:illustration}.
Two spatially separated condensates, described by the two bosonic modes ($c_1$) and ($c_2$), are resonantly pumped by entangled photon pairs generated in a nonlinear medium with, \textit{e.g.}, a quadratic susceptibility ($\chi^{(2)}$), driven by a laser at frequency $\omega$. The condensates are also subject to incoherent pumping from their respective exciton reservoirs and to photon leakage into the environment. The incoherent gain is saturable because the reservoirs are depleted by stimulated scattering into the condensate. This allows the condensate occupations to be stabilized. At the same time, reservoir-induced fluctuations and photon losses tend to degrade the intermode correlations and prevent entanglement.

The stronger the entangled pumping, the larger the pair-induced polariton influx into the two condensates. However, this also shifts the stationary state towards higher condensate occupations. The increased occupation is accompanied by larger total incoherent gain and loss fluxes and, consequently, by stronger associated fluctuations.

Because of this competition, the dependence of condensate entanglement on the entangled-pumping strength is nontrivial and must be determined quantitatively.

While nonresonantly pumped polariton condensates and their optical confinement are well established experimentally, the complete two-condensate setup considered here has not yet been demonstrated. Nevertheless, the coupling of quantum-correlated light to a polaritonic microcavity has already been realized experimentally, including the transfer of correlations from entangled photon pairs to a polariton mode~\cite{cuevasFirstObservationQuantized2018}. This provides experimental support for the key ingredient of our proposal, namely the transfer of photonic quantum correlations to polaritonic modes.

\subsection{Entangled pumping}

Since our aim is a proof-of-principle analysis, we do not explicitly model the coupling of the incident photons through the cavity mirrors. Instead, we model their action on the two polariton modes at the level of an effective coherent two-mode parametric coupling \cite{quantumOPOI,quantumOPOII}
\begin{equation}
\begin{split}\label{eq:Hamiltonian}
H(t) = & \hbar \omega_1 \hat \psi_1^\dagger(t) \hat \psi_1(t) + \hbar \omega_2 \hat \psi_2^\dagger(t) \hat \psi_2(t) \\
& - \hbar \chi \left[ \hat \psi_1^\dagger(t) \hat \psi_2^\dagger(t) e^{-i\omega t} + \hat \psi_1(t) \hat \psi_2(t) e^{i\omega t} \right]
\end{split}
\end{equation}
with $\omega_1$ and $\omega_2$ being the frequencies of the signal and idler modes (which are also the frequencies of the condensate modes) and $\omega = \omega_1+\omega_2$ due to energy conservation. Here $\hat \psi_{1/2}$ are the Heisenberg picture bosonic annihilation operators for the modes $c_{1/2}$, while $\chi$ is the effective strength of the coherent pair drive. Its quantitative relation to the incident photon-pair flux depends on the specific optical coupling scheme; losses and fluctuations associated with this coupling are not included explicitly in the present model.

Hereafter, we set $\hbar=1$. In the coherent state representation, the dynamics of the system is given by the following equation for $j\in \{1, 2\}$:
\begin{align}\label{eq:OPO_evolution}
    id_t\psi_j=\omega_j\psi_j-\chi\ovl\psi_{3-j}e^{-i\omega t}.
\end{align}

\subsection{Exciton pumping and photonic decay}

To describe the polariton condensates above threshold, the coherent dynamics \eqref{eq:OPO_evolution} must be supplemented by gain saturation and decay. At the deterministic mean-field level we adopt the standard model for polariton condensation from \cite{PhysRevB.79.165302}:
\begin{equation}
    \begin{cases}
i\dot{\psi} {=} \left\{ {-}\dfrac{\hbar^2 \nabla^2}{2m} {+} \dfrac{i}{2} \left[ R(n_R) {-} \Gamma \right] {+} g |\psi|^2 {+} 2\tilde{g} n_R \right\} \psi, \\
\dot{n}_R {=} \mathcal P - \gamma_R n_R {-} R(n_R) \, |\psi(\mathbf{r})|^2 {+} D \nabla^2 n_R.
\end{cases}
\end{equation}
Here the classical condensate field $\psi$ for polaritons of mass $m$ is coupled to the excitonic reservoir of density $n_R$ via a scattering rate $R(n_R)$, which in the simplest case may be considered to be a linear function of density $R(n_R)=Rn_R$. The photon leakage through the microcavity mirrors is modeled by the constant $\Gamma$. The parameters $g$ and $\tilde g$ stand for polariton-polariton and exciton-polariton interactions, respectively, while $D$ is the diffusion constant for reservoir excitons. The excitonic reservoir is replenished at a rate $\mathcal P$.

For simplicity, we restrict ourselves to a single-mode description of the condensates. Consistently with the proof-of-principle character of the present treatment, we absorb the mean energy shifts into the mode frequencies $\omega_j$ and neglect the corresponding frequency fluctuations induced by occupation and reservoir-density fluctuations. We also neglect reservoir diffusion because of the large exciton mass and assume sufficiently fast reservoir relaxation, which allows for its adiabatic elimination. As a result, the following equation for the condensate is valid:
\begin{align}\label{eq:mf_Wouters}
	i\dot{\psi} {=} \left\{\omega+ \dfrac{i}{2} \left[ \frac{R}{\gamma_R+R|\psi|^2}\mathcal P  {-} \Gamma \right]  \right\} \psi.
\end{align}
Hereafter we assume that $\psi$ is the classical field for the condensate mode, thus $|\psi|^2$ gives the occupation number, not the density. The pumping constant $\mathcal P$ is the total particle flux to the reservoir. Further details on this reduction can be found in Sec. III of Ref.~\cite{RevModPhys.85.299}.

Equations of this type may be simplified even further while remaining useful for describing open condensates. Namely, one may use just a simple linear gain saturation model as follows~\cite{KeelingBerloff}:
\begin{align}\label{eq:mf_Keeling}
	i\dot{\psi} {=} \left\{\omega+ \dfrac{i}{2} \left[ \mathcal P_0-\alpha |\psi|^2 {-} \Gamma \right]  \right\} \psi.
\end{align}

\subsection{Mean-field description of entangled condensates}
As we will demonstrate below, some of the results of our study do not depend on the specific details of gain saturation. Because of that, we carry out the subsequent analysis for a general model of two identical condensates pumped by entangled photonic pairs:
\begin{align}\label{eq:OPO+polaritons_mf}
    id_t\psi_{j} = \Bigg[\omega_{j}+\frac{i}2\left(P[|\psi_j|^2]-\Gamma[|\psi_j|^2]\right)\Bigg]\psi_{j}-\chi \ovl \psi_{3-j} e^{-i\omega t},
\end{align}
where both the gain $P[|\psi_j|^2]$ and decay $\Gamma[|\psi_j|^2]$ rates may be occupation-dependent. After Madelung transformation $\psi_j=\sqrt{\rho_j}e^{-i\omega_jt+i\theta_j}$, we obtain the following equations for the occupations and phases of the two condensate modes:
\begin{equation}
	\begin{cases}
		d_t \rho_j=\left(P[\rho_j]-\Gamma[\rho_j]\right)\rho_j+2\chi\sqrt{\rho_j\rho_{3-j}}\sin(\theta_1+\theta_2),\\
		d_t \theta_j=\chi \sqrt{\frac{\rho_{3-j}}{\rho_j}}\cos(\theta_1+\theta_2).
	\end{cases}
\end{equation}

Assuming there is a stable equilibrium $P[\rho_j]-\Gamma[\rho_j]=0$ without entangled pumping (otherwise there would be no isolated condensates), we see that the new equilibrium is established at a higher occupation, $\rho_1=\rho_2=\rho_m$:
\begin{align}
	P[\rho_m]-\Gamma[\rho_m]+2\chi=0
\end{align}
since the system favors a fixed value of the phase sum
\begin{align}
\theta_+=\theta_1+\theta_2=\pi/2.
\end{align}
This mean-field picture already captures the two key ingredients of the stationary state: stabilization of the condensate occupation and locking of the phase sum $\theta_+$ by entangled pumping. However, it does not allow one to determine whether the resulting state is quantum mechanically entangled. This requires taking into account fluctuations, which we address in the next section.

\section{Stochastic phase-space model}

To include fluctuations, we extend \eqref{eq:OPO+polaritons_mf} to stochastic Ito equations. A single-mode stochastic description of a nonresonantly pumped polariton condensate may be obtained from a reduced master equation for the condensate mode using the Wigner representation (see Refs.~\cite{Schwendimann2008,Schwendimann2010} for details). In this approach, scattering from the reservoir to the condensate and back is described by the rates $R_{\rm in}$ and $R_{\rm back}$. The net flow enters the drift as their difference, while the noise is determined by their sum.

In the present model, we allow these effective rates to depend on $|\psi_j|^2$, which accounts for gain saturation after the reservoir is adiabatically eliminated. Related microscopic descriptions based on the Schwinger-Keldysh technique can be found in Refs.~\cite{PhysRevB.89.155302, PhysRevB.97.014525}. Photon leakage gives an additional loss rate $\Gamma$ and also contributes to the noise. Thus, adding the incoherent gain and loss processes to the coherent evolution, we obtain
\begin{align}\label{eq:OPO+polaritons}
    id\psi_{j} = \Bigg[\omega_{j}\psi_{j}
    +\frac{i}2\left(P[|\psi_j|^2]-\Gamma[|\psi_j|^2]\right)\psi_{j}\nonumber\\
    -\chi \ovl \psi_{3-j} e^{-i\omega t}\Bigg] dt
    +\sqrt{\frac{1}4\left(Q[|\psi_j|^2]+\Gamma[|\psi_j|^2]\right)} d \bs W_{j}.
\end{align}
Here $d\bs W_j$ are complex Wiener increments with $d\bs W_jd\bs W_j^*=2dt$, and
\begin{align}
    P[|\psi_j|^2]&=R_{\rm in}[|\psi_j|^2]-R_{\rm back}[|\psi_j|^2],\nonumber\\
    Q[|\psi_j|^2]&=R_{\rm in}[|\psi_j|^2]+R_{\rm back}[|\psi_j|^2].
\end{align}

For an equilibrated reservoir at temperature $T_R$ with chemical potential $\mu_R$, the in- and back-scattering rates are related by detailed balance. This gives
\begin{align}\label{eq:eta_definition}
    Q[|\psi_j|^2]=\coth\left(\frac{E-\mu_R}{2kT_R}\right)
    P[|\psi_j|^2]=\eta P[|\psi_j|^2].
\end{align}
Thus, $\eta$ characterizes reservoir noise amplification due to the backflow of particles. Additionally, $\eta=1$ corresponds to the limit of negligible backflow and therefore to the minimum reservoir noise for a given net gain $P$. For exciton-polariton condensates, the energy $E$ corresponds to the energy offset of the lower polariton mode with respect to exciton resonance frequency, as demonstrated below. It depends on the Rabi splitting $\Omega$ and exciton-photon detuning $\Delta_0$ in the following way (see Fig.~2):
\begin{align}
    E=\frac{\Delta_0}{2}-\sqrt{\left(\frac{\Delta_0}{2}\right)^2+\Omega^2},
\end{align}

\begin{figure}[htp]
	\centering
	\includegraphics[]{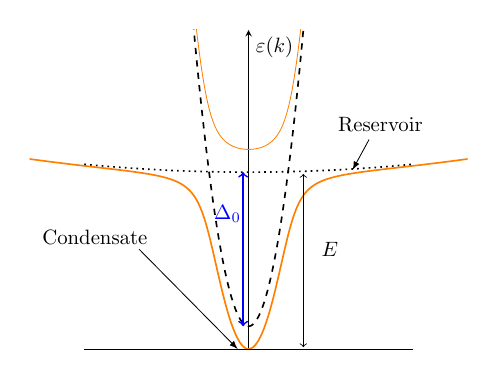}
	\caption{Dispersion curves $\varepsilon(k)$ for the lower and upper polariton modes (orange lines). The dashed and dotted lines depict the bare photon and exciton dispersions, respectively. Their energy offset at zero momentum is denoted by $\Delta_0$. For the lower polariton branch, the energy offset $E$ between the condensate and the reservoir is indicated.}
\end{figure}

We emphasize that the photon-loss drift and noise terms correctly reproduce zero-point fluctuations $\braket{|\psi_j|^2}=1/2$ in the absence of any pumping.

Overall, the model introduced above is a minimal description of the entanglement mechanism. It retains the fluctuations intrinsically associated with incoherent gain and photon loss, while omitting additional implementation-specific effects as well as occupation-dependent frequency fluctuations. Importantly, even within this reduced setting, reaching an entangled stationary state is not guaranteed.

\section{Properties of the stationary state of two coupled condensates}\label{sec:statstate}
Since entanglement criteria for continuous-variable systems are naturally formulated in terms of quadrature covariance matrices~\cite{SimonPeres, Adesso_2007}, we switch to the quadrature representation:
\begin{align}\label{eq:quadratures}
    x_j=\frac{\psi_je^{i\omega_j t}+\ovl\psi_je^{-i\omega_j t}}{2},\nonumber\\
    p_j=\frac{\psi_je^{i\omega_j t}-\ovl\psi_je^{-i\omega_j t}}{2i}
\end{align} to represent the drift term as a gradient flow generated by an effective potential:

\begin{equation}\label{eq:quadrature evolution}
    \begin{cases}
        d x_j&=-\partial_{x_j}Vdt{+}\sqrt{\frac14\left[\eta P[\rho_j]{+}\Gamma[\rho_j]\right]}dW_{xj},\\
        d p_j&=-\partial_{p_j}Vdt{+}\sqrt{\frac14\left[\eta P[\rho_j]{+}\Gamma[\rho_j]\right]}dW_{pj}.
    \end{cases}
\end{equation}
with independent Wiener increments $dW_{xj}$ and $dW_{pj}$. Here, we introduce
\begin{align}\label{eq:general_potential}
    V(x_1,\! p_1,\! x_2,\! p_2)=\frac{G[\rho_1]{+}G[\rho_2]}4-\chi (p_2 x_1 {+} x_2 p_1)
\end{align}
and define $G[\rho]$ by $d_{\rho}G[\rho]=\Gamma[\rho]-P[\rho]$ with $\rho_j=x_j^2{+}p_j^2$.

Hereafter we restrict ourselves to the states with high mean occupations $\braket{\rho_j}\gg1$ of the condensate modes. This allows us to treat the noise variance terms as constants. With that assumption, the stationary Wigner distribution may be explicitly derived (see \cite{gardiner_handbook_1983_ch5}):
\begin{equation}\label{eq:Wigner_stationary}
    W(x_1, p_1, x_2, p_2)\sim \exp\left(-\frac{2V(x_1, p_1, x_2, p_2)}{\frac14\left(\eta P[\braket{\rho_j}]+\Gamma[\braket{\rho_j}]\right)}\right).
\end{equation}
The additive constant in $G[\rho]$ is absorbed into the normalization of the Wigner function. And due to the symmetry of the problem, $\braket{\rho_1}=\braket{\rho_2}$.

Note that without a flux of entangled pairs, $\chi=0$, the distribution factorizes:
\begin{align}\label{eq:Wigner_factorized}
	W(x_1, p_1, x_2, p_2)=W^{(1)}(x_1,p_1)W^{(1)}(x_2, p_2)
\end{align}
with 
\begin{align}\label{eq:Wigner_factor}
	W^{(1)}(x_i, p_i)\sim \exp\left(-\frac{G[x_i^2+p_i^2]}{\frac12\left(\eta P[\braket{x_i^2+p_i^2}]+\Gamma[\braket{x_i^2+p_i^2}]\right)}\right).
\end{align}
Above the condensation threshold $G[\rho]$ is characterized by a well-defined minimum that leads to independent ring-shaped distributions with reduced intensity fluctuations and a uniform phase distribution. It is a typical stationary state for two-dimensional (2D) symmetry-broken systems, such as the single-mode laser \cite{gardiner_quantum_2010_laser} and atomic Bose-Einstein condensates \cite{Stoof1999_63-64}. In terms of amplitude fluctuation statistics, they are barely distinguishable from coherent states.

\begin{figure}[htp]
    \centering
    \includegraphics[width=0.8\linewidth]{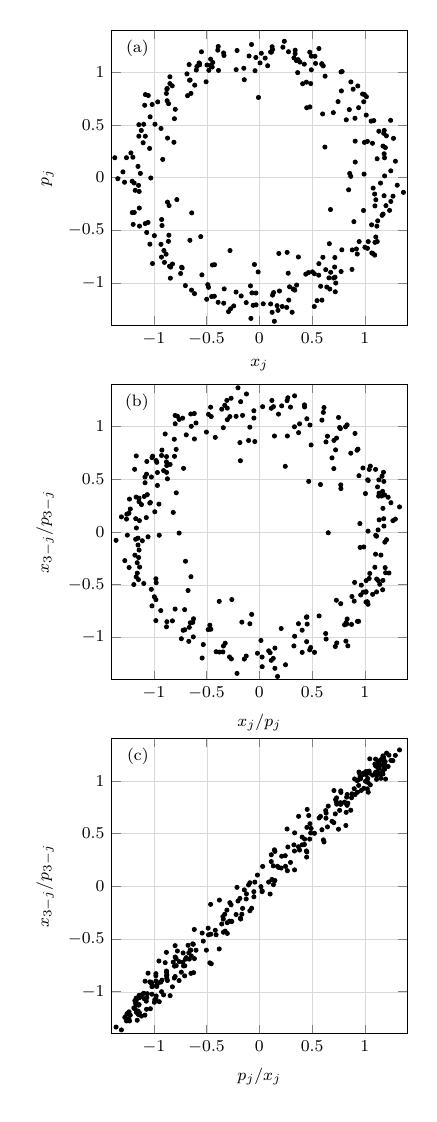}
    \caption{Wigner quasiprobability distribution illustrated for the state of type \eqref{eq:Wigner_stationary}. The marginal distributions for (a) $p_j(x_j)$; (b) $x_j(x_{3-j})$ (the distribution for $p_j(p_{3-j})$ has the same form); (c) $x_j(p_{3-j})$ [the distribution for $p_j(x_{3-j})$ has the same form] are shown. The Hamiltonian \eqref{eq:Hamiltonian} generates squeezing in quadratures $x_j-p_{3-j}$, which results in deformation of the distributions in panel (c). These illustrative pictures are generated by simulating stochastic evolution governed by \eqref{eq:quadrature evolution} with the simplest quadratic model for $G[\rho]=\rho^2-2\rho$, $\eta=1$, $\chi=0.4$.}
    \label{fig:distribution}
\end{figure}

In both states \eqref{eq:Wigner_stationary} and \eqref{eq:Wigner_factorized} the marginal distributions in the $(x_j, p_j)$ planes are qualitatively the same. However, the quadratures $(x_1, p_2)$ and $(x_2, p_1)$ are squeezed in the state \eqref{eq:Wigner_stationary} as schematically demonstrated in Fig.~\ref{fig:distribution}.

We now utilize the Madelung transformation $x_j=\sqrt{\rho_j}\cos(\theta_j)$, $p_j=\sqrt{\rho_j}\sin(\theta_j)$ to express the potential as follows:
\begin{align}\label{eq:potential}
    V(\rho_1, \rho_2,\theta_+)=\frac14\left(G(\rho_1)+G(\rho_2)\right)-\chi \sqrt{\rho_1\rho_2} \sin(\theta_+)
\end{align}
with $\theta_+=\theta_1+\theta_2$. 

The phase-space structure of Eq.~\eqref{eq:Wigner_stationary} along with Eq.~\eqref{eq:potential} has a close counterpart in the exact steady-state solution of the nondegenerate parametric oscillator~\cite{Kheruntsyan2000}. In that case, the joint Wigner function also depends on the product of the two mode amplitudes and is therefore sensitive to their phase sum, whereas the marginal Wigner functions for individual modes are rotationally symmetric. The same residual phase symmetry is present here: the stationary distribution depends on $\theta_+=\theta_1+\theta_2$, while $\theta_-=\theta_1-\theta_2$ remains free. The underlying dissipative mechanisms are, however, different: in the present system each mode is independently sustained by saturable incoherent gain and is subject to the associated fluctuations.

The minimum of this potential is given by $\rho_1=\rho_2=\rho_m$, $\theta_1+\theta_2=\pi/2$ with
\begin{align}\label{eq:equilibrium_rho_cond}
    P[\rho_m]+2\chi=\Gamma[\rho_m].
\end{align}
In the weak-noise limit, \textit{i.e.,} $\rho_m\gg\sqrt{{\rm Var}({\rho_j})}$, we may use a Gaussian expansion for the potential \eqref{eq:potential}, substitute it into the stationary Wigner function \eqref{eq:Wigner_stationary} and integrate over the phase difference $\theta_-=\theta_1-\theta_2$:
    \begin{widetext}
\begin{align}\label{eq:gaussian_potential}
    W(\rho_1, \rho_2, \theta_+)=\mathcal N\exp\!\left[-\frac12\begin{pmatrix}
        \rho_1{-}\rho_m& \rho_2{-}\rho_m
    \end{pmatrix}\begin{pmatrix}
        C_{11}&C_{12}\\
        C_{12}&C_{22}
    \end{pmatrix}^{-1}\!\!\begin{pmatrix}
        \rho_1{-}\rho_m\\ \rho_2{-}\rho_m
    \end{pmatrix}-\frac1{2C_{\theta}}\left(\theta_+-\frac\pi2\right)^2\right].
\end{align}
\end{widetext}
Here ($\delta \theta = \theta_+-\pi/2$)
\begin{align}
    C&=\braket{\delta\rho_j^2}=\frac{((1{+}\eta)\Gamma[\rho_m]{-}2\eta\chi)(G'[\rho_m] \rho_m{+}\chi)}{2G'[\rho_m](G'[\rho_m]\rho_m+2\chi)},\label{eq:C}\\
    C_{12}&=\braket{\delta\rho_1\delta\rho_2}=\frac{(1{+}\eta)\Gamma[\rho_m]{-}2\eta\chi}{2G'[\rho_m](G'[\rho_m]\rho_m+2\chi)}\chi,\label{eq:C12}\\
    C_\theta&=\braket{\delta\theta^2}=\frac{(1{+}\eta)\Gamma[\rho_m]{-}2\eta\chi}{8\chi\rho_m}.\label{eq:Cth}
\end{align}

We stress that the Gaussian approximation in Eq.~\eqref{eq:gaussian_potential} does not make the state Gaussian in the canonical quadratures. This differs from the linearized Gaussian treatments commonly used in exciton-optomechanical entanglement schemes~\cite{Zuo2024,EsQars2025}, where quantum fluctuations are linearized around a coherently driven steady state and the resulting Gaussian state is fully characterized by its covariance matrix.

When transforming back to quadratures, one can derive the following expression for the covariance matrix $\sigma$ (see Appendix \ref{app:covariance_matrix}):
\begin{align}\label{eq:covar_matrix}
    \sigma &=\begin{pmatrix}
\langle \Delta x_1^2 \rangle & \langle \Delta x_1 \Delta p_1 \rangle & \langle \Delta x_1 \Delta x_2 \rangle & \langle \Delta x_1 \Delta p_2 \rangle \\
\langle \Delta p_1 \Delta x_1 \rangle & \langle \Delta p_1^2 \rangle & \langle \Delta p_1 \Delta x_2 \rangle & \langle \Delta p_1 \Delta p_2 \rangle \\
\langle \Delta x_2 \Delta x_1 \rangle & \langle \Delta x_2 \Delta p_1 \rangle & \langle \Delta x_2^2 \rangle & \langle \Delta x_2 \Delta p_2 \rangle \\
\langle \Delta p_2 \Delta x_1 \rangle & \langle \Delta p_2 \Delta p_1 \rangle & \langle \Delta p_2 \Delta x_2 \rangle & \langle \Delta p_2^2 \rangle
\end{pmatrix}\nonumber\\
&= \frac{\rho_m}2\begin{pmatrix}
        1&0&0&\xi\\
        0&1&\xi&0\\
        0&\xi&1&0\\
        \xi&0&0&1
    \end{pmatrix},
\end{align}
where
\begin{align*}
	\Delta x_j&=x_j-\braket{x_j},\\
	\Delta p_j&=p_j-\braket{p_j}
\end{align*}
and 
\begin{align*}
    \xi = 1-\frac{C-C_{12}}{4\rho_m^2}-\frac{C_\theta}{2}
\end{align*}
is the effective squeezing degree of the stationary state. It is helpful to see how the reduced values of the variance in $\theta_+$ and the difference $C-C_{12}$ lead to the squeezing effect shown in Fig.~\ref{fig:distribution}(c). For example, we may rotate the coordinate axes in the $(x_1, p_2)$ plane:
\begin{align}
	q_{\pm}=\frac{x_1\pm p_2}{\sqrt{2}}.
\end{align}
The variance ratio $K$ along the two axes may serve to quantify the distribution anisotropy. As shown in detail in Appendix \ref{app:covariance_matrix}, we derive:
\begin{align}\label{eq:anisotropy}
	K=\frac{\braket{q_-^2}}{\braket{q_+^2}}=\frac{\frac12\left[\frac{C-C_{12}}{4\rho_m^2}+\frac{C_\theta}{2}\right]}{1-\frac12\left[\frac{C-C_{12}}{4\rho_m^2}+\frac{C_\theta}{2}\right]}=\frac{1-\xi}{1+\xi}.
\end{align} 
\section{Test for entanglement}\label{sec:test}
The stationary state \eqref{eq:Wigner_stationary} [and also its weak-noise approximation \eqref{eq:gaussian_potential}] of condensates exposed to external pumping by entangled photon pairs is a highly non-Gaussian state in the canonical quadratures, as is typical for the ring-shaped states (see \cite{Allevi:13}). For Gaussian states, the first two moments provide a complete characterization of the state, which enables the use of a broad range of covariance-matrix-based entanglement criteria~\cite{Adesso_2007}. This is no longer the case for the non-Gaussian states, including the one considered here. However, we can still use the covariance-matrix form of the Peres-Horodecki, or positive partial transpose (PPT), criterion as a necessary condition for separability of an arbitrary bipartite state. Even though satisfaction of the separability condition is inconclusive (since entanglement not captured by the second moments may remain undetected), the violation of the separability criterion is enough to guarantee entanglement. This is sufficient for the goal of this paper, which is to demonstrate the principal possibility of entanglement.
\subsection{General form of flux saturation}
Having derived the correlation matrix for the stationary state, we may utilize the adaptation of the PPT criteria for continuous variable states by Simon~\cite{SimonPeres}. Having identified the blocks of the covariance matrix:
\begin{align}\label{eq:peres_covar}
    2\sigma =\begin{pmatrix}
        A&C\\
        C^T&B
    \end{pmatrix},
\end{align}
we should check the following inequality:
\begin{align}
    f_{\rm PT}&{=}\det A \det B {+} \!\!\left( \frac{1}{4} {-} |\det C| \right)^2\!\!\! 
{-} \operatorname{tr}(AJCJBJC^TJ) \nonumber\\
&{-} \frac{1}{4} \left( \det A {+} \det B \right)<0
\end{align}
as a sufficient (yet not necessary) condition for quantum entanglement. Here
\begin{align*}
J=\begin{pmatrix}
    0&1\\-1&0
\end{pmatrix}.    
\end{align*}
The factor of $2$ on the left-hand side of Eq.~\eqref{eq:peres_covar} is due to the additional factor of $\sqrt{2}$ in Eq.~\eqref{eq:quadratures} as compared to the definitions of canonical coordinate and momentum.

We divide the matrix \eqref{eq:covar_matrix} into four blocks and calculate the function $f_{\rm PT}$. For the PPT entanglement condition to be satisfied, the squeezing of the state should eventually exceed a threshold value
\begin{align}\label{eq:squeezing_inequality}
    \xi>1-\frac{1}{2\rho_m}.
\end{align}

Careful inspection of Eqs.~\eqref{eq:C} to \eqref{eq:Cth} shows that $\xi$ can be expressed as a function of only two combinations of variables. The first is the relative entangled pumping strength with respect to the decay rate. At the symmetric stationary point, the pair drive contributes an influx $J_{\rm pair}=2\chi\rho_m$ into each condensate, whereas the decay flux is $J_{\rm loss}=\Gamma[\rho_m]\rho_m$. Therefore,
\begin{align}
    \zeta =\frac{J_{\rm pair}}{J_{\rm loss}} = \frac{2\chi}{\Gamma[\rho_m]}.
\end{align}
The second is the flux elasticity $\kappa$ with respect to changes in the condensate occupation. It quantifies the change in the net loss rate, normalized by the decay rate $\Gamma$, induced by a relative change in the condensate occupation:
\begin{align}
    \frac{\delta[\Gamma(\rho)-P(\rho)]}{\Gamma(\rho_m)}\simeq\kappa\frac{\delta\rho}{\rho_m}.
\end{align}
Here $\rho = \rho_m+\delta \rho$. Using the function $G[\rho]$, introduced in \eqref{eq:general_potential}, we express
\begin{align}
    \kappa = \frac{G'[\rho_m]\rho_m}{\Gamma[\rho_m]}.
\end{align}
Note that $\kappa>0$ is a necessary condition for the stability of the condensate.

\begin{figure}
    \centering
    \includegraphics[width=\linewidth]{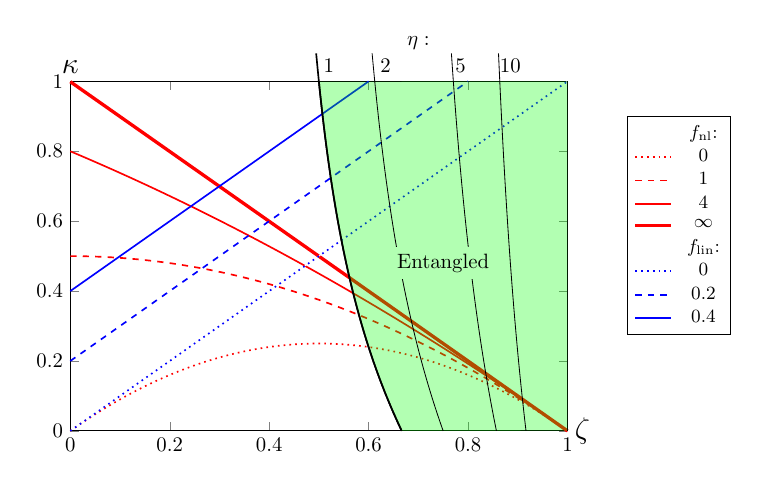}
    \caption{Phase diagram for polariton entanglement. The region of entanglement  $\zeta>\zeta^{\eta = 1}_{\rm crit}(\kappa)$\ is shaded in green and limited by the lines $\zeta^{\eta}_{\rm crit}(\kappa)$ for $\eta>1$ (the higher is the reservoir effective temperature $T_R$, the higher is $\eta$). Red lines correspond to the dependence $\kappa(\zeta)$ within the model \eqref{eq:saturation_model}, blue ones are the ones given by the model \eqref{eq:linear_saturation_model}, $f$ is the normalized excess pump intensity as defined by \eqref{eq:excess}.}
    \label{fig:entanglement_boundary}
\end{figure}

The condition \eqref{eq:squeezing_inequality} for the buildup of entanglement can be expressed as
\begin{align*}  
    2(2+\eta)\zeta^2+\left\{(4+\eta)\kappa(\zeta)-2(1+\eta)\right\}\zeta-(1+\eta)\kappa(\zeta)>0,
\end{align*}
which gives the following equation for the critical value:
\begin{widetext}
\begin{align}
\zeta_{\rm crit}=\frac{2 \eta +\sqrt{4 (\eta +1)^2+(\eta +4)^2
   \kappa^2(\zeta_{\rm crit})+4 \eta  (\eta +1) \kappa(\zeta_{\rm crit})}-(\eta +4) \kappa(\zeta_{\rm crit})+2}{4
   (\eta +2)}.
\end{align}
\end{widetext}
Clearly, the lowest critical value $\zeta_{\rm crit}$ corresponds to the absence of backflow from the excitonic reservoir ($\eta=1$):
\begin{align}\label{eq:critpump_eta1}
    \zeta_{\rm crit}^{\eta=1}=\frac{1}{12} \left(\sqrt{25 \kappa^2+8 \kappa+16}-5 \kappa+4\right)\in \left[\frac{2}{5};\frac{2}{3}\right].
\end{align}
The entanglement region defined by $\zeta>\zeta^{\eta =1}_{\rm crit}$ is shown as a green area in Figure \ref{fig:entanglement_boundary} which is bounded by the lines $\zeta_{\rm crit}(\kappa)$ corresponding to the different values of the reservoir temperature effectively accounted for by $\eta>1$, see \eqref{eq:eta_definition}.

The very existence of the region above $\zeta_{\rm crit}$ is an important result of this work. This means that, even in the highly nonequilibrium system such as the polaritonic one, the fundamental fluctuation-dissipation relations do not prevent the  condensates from being entangled at sufficiently high $\chi$.

\subsection{Back to polaritons}

As discussed in Sec.~\ref{sec:model}, for polaritons the decay rate is mainly due to photon leakage and is weakly dependent on the density of the condensate $\Gamma\approx\rm const$. Thus, the exact values of $\zeta_{\rm crit}$ depend on the saturation model used to describe the flow of particles from the excitonic reservoir. The gain saturation terms are given by expressions
\begin{align}\label{eq:saturation_model}
    P_{\rm nlin}[\rho]&=\frac{PR}{\gamma_R+R\rho}
\end{align}
and
\begin{align}\label{eq:linear_saturation_model}
    P_{\rm lin}[\rho]&=P_0-\alpha\rho
\end{align}
for the models \eqref{eq:mf_Wouters} and \eqref{eq:mf_Keeling}, respectively.

Notably, for both models, the relation between $\zeta$ and $\kappa$ is characterized by only one additional parameter
\begin{align}\label{eq:excess}
f=\frac{P}{P_{\rm th}}-1,
\end{align}
which is the normalized excess pump intensity above the threshold value $P_{\rm th}$ (defined as the one corresponding to the vanishing condensate $\rho\to 0$). We derive the dependence of the flux elasticity $\kappa$ on the entangled-pumping intensity $\zeta$:
\begin{align*}
    \kappa(\zeta)&=\frac{[1 -\zeta] [\zeta +f]}{1+f}
\end{align*}
for the saturation model \eqref{eq:saturation_model} and
\begin{align}\label{eq:kappa_linear}
    \kappa(\zeta) = f+\zeta
\end{align}
for the linear saturation model \eqref{eq:linear_saturation_model}.
A series of curves is depicted in Figure \ref{fig:entanglement_boundary} for both models showing the dependence $\zeta(\kappa)$ for various values of $f$. Intersecting the corresponding curve with the entanglement boundary for a particular value of $\eta$ (dependent, \textit{e.g.}, on reservoir temperature, exciton-photon detuning, reservoir chemical potential) gives the minimum normalized entangled-pumping flux $\zeta$ required to reach an entangled stationary state. The significant difference between the lines for the two models at high values of $\zeta$ is due to the artificial nature of the linear saturation model, which technically allows for negative $P_{\rm lin}[\rho]$, \textit{i.e.} reversed particle flow from the condensate to the reservoir.

We expect a real polaritonic system to lose stability ($\kappa<0$) at a sufficiently high intensity of pumping with entangled pairs. More specifically, the equilibrium condition \eqref{eq:equilibrium_rho_cond} suggests that for $2\chi>\Gamma[\rho_m]$ the pumping term $P[\rho_m]$ should be negative, which is not feasible in polaritonic systems. That is why we show only the region $\zeta\in [0;1]$ in Figure \ref{fig:entanglement_boundary}. Note that $\zeta=1$ is exactly the point where the model \eqref{eq:saturation_model} predicts unstable density growth.

\section{Entanglement lifetime}\label{sec:lifetime}
Given that it is possible to reach an entangled state, we are interested in its lifetime. In this section, we will focus on the simplest linear saturation model with $P(\rho)=P_0-\alpha\rho$ and a constant decay rate $\Gamma$, which allows for analytical solutions. We consider the following ``driving protocol'' for the condensates under consideration:
\begin{align}
    \chi(t)=\Theta(-t)\chi_0,
\end{align}
\textit{i.e.}, a sudden switch-off. Using Ito's lemma, one may derive the following equations for $t>0$ from \eqref{eq:quadrature evolution}-\eqref{eq:general_potential} with $\chi=0$ (see Appendix \ref{app:B} and, \textit{e.g.} , \cite{gardiner_handbook_1983_ch4}):
\begin{widetext}
\begin{align}
    d\rho_j &= \left[ \left(P_0-\alpha\rho_j\right)\left(\rho_j+\frac12\right)-\Gamma \left(\rho_j - \frac12\right)\right] dt + \sqrt{\left(P_0-\alpha\rho_j+\Gamma\right) \rho_j} d W_j(t),\label{eq:radial_Ito}\\
    d\theta_j &=\sqrt{\frac{1}4\frac{P_0-\alpha\braket{\rho_j}+\Gamma}{\braket{\rho_j}}}d\Omega_j\label{eq:angular_Ito}
\end{align}
\end{widetext}
with independent Wiener increments $dW_j$ and $d\Omega_j$.
Here we also use the assumption $\braket{\rho_j}\gg\sqrt{\rm Var({\rho_j})}$ and set $\eta$ to unity (no backflow from the exciton reservoir) for simplicity.
Tracing the evolution of the Wigner function in the weak-noise limit, where the Gaussian approximation is valid in terms of $\rho_j$ and $\theta_+$, we may use the mean values $\rho_m(t)=\braket{\rho_j}(t)$ when calculating the noise variance and apply Wick's or Isserlis' theorem to derive (see Appendix ~\ref{app:B} for details):
\begin{equation}
\begin{cases}
    d_t\rho_m&=(P_0-\alpha \rho_m-\Gamma)\rho_m\\
    d_tC&=2(P_0-2\alpha \rho_m-\Gamma)C+\left(P_0-\alpha \rho_m+\Gamma\right)\rho_m\\
    d_tC_{12}&=2(P_0-2\alpha \rho_m-\Gamma)C_{12}.
\end{cases}
\end{equation}

Here we assume that the mean occupations evolve independently and act as "external" parameters for $C=\braket{\delta\rho_j^2}(t)$ and $C_{12}=\braket{\delta\rho_1\delta\rho_2}(t)$. Under these approximations, it is straightforward to obtain an analytical solution. Specifically, if we use $\tau=\Gamma t$ as a time variable,
\begin{align}\label{eq:m(t)}
    \rho_m(\tau)=\frac{f\Gamma}{\alpha}\frac{f+\zeta_0}{f+\zeta_0(1-e^{-f\tau})}
\end{align}
and hence, using Eq.~\eqref{eq:angular_Ito},
\begin{align}\label{eq:angular_diffusion}
    C_\theta(\tau)=C_\theta(0)+\frac12\int_0^\tau \left(\frac{f+2}{\rho_m(z)}-\frac{\alpha}{\Gamma}\right)dz.
\end{align}
Having the analytical expressions for means and variances (see Appendix ~\ref{app:B} for complete expressions), one can trace the evolution of the squeezing degree $\xi(\tau)$ to see when it crosses the threshold value $1-1/(2\rho_m(\tau))$ as suggested by Eq.~\eqref{eq:squeezing_inequality} (the structure of the covariance matrix does not change during evolution). This defines the entanglement lifetime $\tau_d$ within the present criterion, which characterizes the time interval $[0;\tau_d]$ during which the system is guaranteed to remain entangled. In the general case, one can find $\tau_d$ only numerically. A simple analytical expression serves as an upper bound for $\tau_d$ (see Appendix~\ref{app:C}):
\begin{align}
\tau_d < \tau_d^{\rm ub}
=\frac{1}{2}\left(\frac{5}{2}-\frac{1}{\zeta_0}\right)
+\frac{\zeta_0}{4f}\frac{f+2}{f+\zeta_0}.
\end{align}
Numerical evaluation using the complete analytical solutions from Appendix~\ref{app:B} shows that, for fixed $\zeta_0$, the bound becomes tighter with increasing $f$ and is asymptotically exact for $f(\zeta_0-2/5)\gg 1$. In this regime, the exact value of $\tau_d$ is asymptotically given by
\begin{align}
\tau_d(\zeta_0)\simeq\frac{5}{4\zeta_0}\left[\zeta_0-\frac{2}{5}\right].
\end{align}
This expression shows how the entanglement lifetime depends on the relative entangled pumping strength at $t<0$. Comparing this expression with Eq.~\eqref{eq:m(t)}, one may see that the relaxation timescale for the condensate density is $\sim[f\Gamma]^{-1}$, while the entanglement lifetime is of the order of $\Gamma^{-1}$.

\section{Discussion and conclusion}\label{sec:disc}
This work is intended to demonstrate that two spatially separated polariton condensates can be entangled despite their noisy evolution due to coupling to the exciton reservoirs and photon leakage. We provide estimates for the required values of the control parameters.

The present proof-of-principle results should be viewed as a baseline for further, more detailed analysis assessing, in a specific experimental realization, how additional implementation- and material-specific effects modify the robustness of entanglement.

In addition, we demonstrated that, within the reduced model considered here, the entanglement does not survive much longer than the photon lifetime in the cavity. Importantly, this estimate is obtained after neglecting setup-specific decoherence mechanisms and taking the reservoir noise at its minimum value, $\eta=1$, allowed by the fluctuation-dissipation relation. Thus, the remaining fluctuations represent an intrinsic source of noise that cannot be consistently excluded from the present dissipative description. In this sense, the timescale $\sim\Gamma^{-1}$ sets an optimistic limiting scale for the entanglement lifetime within the passive evolution considered here. Additional relaxation, dephasing, and other implementation-specific effects are expected to shorten this lifetime. In agreement with the intuitive expectation, increasing the entanglement lifetime within this scheme requires increasing the photon lifetime in the cavity.

At the same time, the entanglement and the quadrature anisotropy decay on very different timescales. Analysis of the expression \eqref{eq:anisotropy} along with evolution Eqs.~\eqref{eq:radial_Ito} and \eqref{eq:angular_Ito} implies that after the entangled-pair pumping is turned off, the amplitude correlations relax at the scale of $\sim [f\Gamma]^{-1}$, the entanglement vanishes at the scale of $\sim \Gamma^{-1}$, although the ratio $[C-C_{12}]/\rho_m^2$ remains small (as a consequence of the weak noise approximation for condensates above threshold). Importantly, the anisotropy relaxation timescale is mainly defined by the growth of $C_{\theta}$. Given the factor of $1/\rho_m$ in Eq.~\eqref{eq:angular_diffusion}, the time estimate for $C_{\theta}$ to become of the order of unity is $\rho_m \Gamma^{-1}$, which is much longer (for typical experimental sample sizes and densities, condensate occupation $\rho_m$ is of the order of $10^2-10^3$) than the density relaxation time. The entanglement criterion is much more demanding in terms of the degree of squeezing of the $(x_j, p_{3-j})$ distribution, as shown by the criterion \eqref{eq:squeezing_inequality}. However, analyzing Eq. \eqref{eq:anisotropy} along with expressions from Appendix \ref{app:C}, one may see that the quadrature anisotropy itself persists for a much longer time than the entanglement. Correlations persisting beyond entanglement can remain relevant for parameter-estimation tasks, as known for correlated separable continuous-variable states~\cite{Adesso2014GIP}. Whether the long-lived correlations found here can provide a metrological advantage requires a separate analysis beyond the scope of present work.

It is important to note that in the considered toy model the two polariton condensates are assumed to be spatially separated, and, apart from the common entangled-pair drive, do not interact directly. No interactions between polaritons belonging to different condensates or excitons belonging to different reservoirs are taken into account. In contrast, quantum processors based on polariton qubits would require controllable coupling between trapped polariton condensates~\cite{Barrat2024-af}. Another important remark concerns the estimate of the entanglement lifetime given in this work. In a recent experimental study of a qubit based on a trapped exciton-polariton condensate~\cite{opt6040053}, the relaxation and coherence times $T_1$ and $T_2$ are found to exceed the single-polariton lifetime by about two orders of magnitude. These times characterize a different degree of freedom and therefore cannot be directly identified with the entanglement lifetime estimated here. Nevertheless, this result motivates studying the coherence and entanglement dynamics of pairs of such condensates, in particular when they are initially driven into an entangled state as suggested in our work.
\newpage
\bibliographystyle{unsrt}
\bibliography{references.bib}
\clearpage
\appendix
\appendixpage
\addappheadtotoc
\section{Covariance matrix}\label{app:covariance_matrix}
In this appendix, we evaluate the covariance matrix elements, using the Gaussian approximation in $\rho_1$, $\rho_2$, and $\theta_{+}$. Since the stationary distribution depends on the phases only through their sum, the marginal distribution of either individual phase is uniform. Moreover, the distribution of amplitude and phase fluctuations factorizes, see Eq.~\eqref{eq:gaussian_potential}. We start from $\braket{x_1^2}$:
\begin{align}
\braket{x_1^2}&=\braket{\rho_1\cos^2(\theta_1)}=\braket{(\rho_m+\delta \rho_1)}\braket{\cos^2(\theta_1)}\nonumber\\
&=\frac12\left(\rho_m+\braket{\delta \rho_1}\right)=\frac{\rho_m}2.
\end{align}
Similar arguments lead to $\braket{x_{1/2}^2}=\braket{p_{1/2}^2}=\rho_m/2$.
Clearly, $\braket{x_1p_1}=\braket{x_2p_2}=0$ since $\braket{\sin(\theta_{1/2})\cos(\theta_{1/2})}=0$. Next, using $\theta_1=\pi/2-(\theta_2-\delta\theta)$:
\begin{widetext}
\begin{align}
    \braket{x_1x_2}&=\braket{\sqrt{\rho_1\rho_2}\cos(\theta_1)\cos(\theta_2)}=\Bigg\langle\rho_m\left(1{+}\frac{\delta\rho_2{+}\delta\rho_1}{2\rho_m}{+}\frac{\delta\rho_1\delta\rho_2}{2\rho_m^2}{-}\frac18\left(\frac{\delta\rho_2{+}\delta\rho_1}{\rho_m}\right)^2\right)\cos(\theta_2)\sin(\theta_2-\delta\theta)\Bigg\rangle=\nonumber\\
    &=\Bigg\langle\rho_m\left(1{-}\frac{(\delta\rho_1{-}\delta\rho_2)^2}{8\rho_m^2}\right)\cos(\theta_2)\sin(\theta_2)\left[1-\frac{\delta\theta^2}2\right]\Bigg\rangle=0.
\end{align}
In the same fashion $\braket{p_1p_2}=0$. The crucial non-zero terms are $\braket{x_1p_2}=\braket{x_2p_1}$. They are evaluated as follows:
\begin{align}
\braket{x_2p_1}&=\braket{\sqrt{\rho_1\rho_2}\sin(\theta_1)\cos(\theta_2)}=\Bigg\langle\rho_m\left(1{+}\frac{\delta\rho_2{+}\delta\rho_1}{2\rho_m}{+}\frac{\delta\rho_1\delta\rho_2}{2\rho_m^2}{-}\frac18\left(\frac{\delta\rho_2{+}\delta\rho_1}{\rho_m}\right)^2\right)\cos(\theta_2)\cos(\theta_2-\delta\theta)\Bigg\rangle=\nonumber\\
    &=\Bigg\langle\rho_m\left(1{-}\frac{\delta\rho_2^2{-}2\delta\rho_2\delta\rho_1{+}\delta\rho_1^2}{8\rho_m^2}\right)\cos^2(\theta_2)\left[1-\frac{\delta\theta^2}2\right]\Bigg\rangle=\frac12\rho_m\Bigg\langle\left(1{-}\frac{\delta\rho_2^2{-}2\delta\rho_2\delta\rho_1{+}\delta\rho_1^2}{8\rho_m^2}\right)\left[1-\frac{\delta\theta^2}2\right]\Bigg\rangle=\nonumber\\
    &=\frac{\rho_m}2\Bigg\langle1{-}\frac{\delta\rho_2^2{-}2\delta\rho_2\delta\rho_1{+}\delta\rho_1^2}{8\rho_m^2}-\frac{\delta\theta^2}2{+}\frac{\delta\rho_2^2{-}2\delta\rho_2\delta\rho_1{+}\delta\rho_1^2}{8\rho_m^2}\frac{\delta\theta^2}2\Bigg\rangle= \frac{\rho_m}2\left[1{-}\frac{\braket{\delta\rho_{2/1}^2}{-}\braket{\delta\rho_2\delta\rho_1}}{4\rho_m^2}{-}\frac{\braket{\delta\theta^2}}{2}\right].
\end{align}
\end{widetext}
Given that all the first-order quadrature averages are zero, these expressions justify the structure of the covariance matrix presented in the main text. In addition to these calculations, it is instructive to express the variances in rotated quadratures, which are used in the main text, \textit{e.g.,} $q_{\pm}=\frac{x_1\pm p_2}{\sqrt{2}}$:
\begin{align}
	\braket{q_{\pm}^2}&=\frac{\braket{x_1^2}+\braket{p_2^2}}2\pm\braket{x_1p_2}\nonumber\\
	&=\frac{\rho_m}2\left[1\pm1\mp\left(\frac{\braket{\delta\rho^2_{1/2}}-\braket{\delta\rho_2\delta\rho_1}}{4\rho_m^2}{+}\frac{\braket{\delta\theta^2}}{2}\right)\right].
\end{align}
\section{Evolution equations for the means and cumulants}\label{app:B}
In this appendix, we provide derivations of the results of Sec.~\ref{sec:lifetime}. We start from Eq.~\eqref{eq:quadrature evolution} assuming a constant decay rate $\Gamma$, the linear pumping saturation model \eqref{eq:linear_saturation_model}, $\chi=0$, and $\eta=1$:
\begin{align}
\begin{cases}
    dx_j=\frac12\left[\left(P_0{-}\alpha \rho_j{-}\Gamma\right)x_j\right]dt+\sqrt{\frac14\left(P_0{-}\alpha \rho_j{+}\Gamma\right)}dW_{xj},\\
    dp_j=\frac12\left[\left(P_0{-}\alpha \rho_j{-}\Gamma\right)p_j\right]dt+\sqrt{\frac14\left(P_0{-}\alpha \rho_j{+}\Gamma\right)}dW_{pj}.
\end{cases}
\end{align}
We then apply the inverse Madelung transformation $\rho_j=x_j^2+p_j^2$, $\theta_j=\arctan(p_j/x_j)$ and use Ito's lemma to derive the following equations:
\begin{widetext}
\begin{align}
    d\rho_j &=2x_jdx_j {+} 2p_jdp_j {+} (dx_j)^2 {+} (dp_j)^2= \left[ \left(P_0{-}\alpha\rho_j\right)\left(\rho_j{+}\frac12\right)-\Gamma \left(\rho_j {-} \frac12\right)\right] dt + \sqrt{\left(P_0-\alpha\rho_j+\Gamma\right) \rho_j} \, d W_j(t),\label{eq:app_radial_Ito}\\
    d\theta_j &=\frac{\partial\theta_j}{\partial x_j} dx_j+\frac{\partial\theta_j}{\partial p_j} dp_j
+\tfrac12\left(\frac{\partial^2\theta_j}{\partial x_j^2}(dx_j)^2+2\frac{\partial^2\theta_j}{\partial x_j\partial p_j}dx_jdp_j+\frac{\partial^2\theta_j}{\partial p_j^2}(dp_j)^2\right)=\sqrt{\frac{1}4\frac{P_0-\alpha\rho_j+\Gamma}{\rho_j}}d\Omega_j(t)\label{eq:app_angular_Ito}.
\end{align}
Here we introduce the angular and radial Wiener increments
\begin{align}
    \begin{pmatrix}
dW_j\\
d\Omega_j
\end{pmatrix}
=\begin{pmatrix}
\cos\theta_j & \sin\theta_j\\
-\sin\theta_j & \cos\theta_j
\end{pmatrix}
\begin{pmatrix}
dW_{xj}\\
dW_{pj}
\end{pmatrix}
\end{align}
that are independent because the transformation is orthogonal. The rate equations for the mean values are given by:
\begin{align}
    d_t\braket{\rho_j}=\left(P_0-\Gamma-\frac{\alpha}2\right)\braket{\rho_j}-\alpha\braket{\rho_j}^2-\alpha C_j+\frac{P_0+\Gamma}{2}.
\end{align}
For the second moments $C_j=\braket{\delta \rho_j^2}$ and $C_{12}=\braket{\delta\rho_1\delta\rho_2}$, we apply Ito's lemma once more and substitute Eq. \eqref{eq:app_radial_Ito}:
\begin{align}
&d_tC_j=\frac{2\braket{\rho_j d\rho_j}+\braket{(d\rho_j)^2}-2\braket{\rho_j}d\braket{\rho_j}}{dt}=2\left\langle\left[ \left(P_0-\alpha\rho_j\right)\left(\rho_j+\frac12\right)-\Gamma \left(\rho_j - \frac12\right)\right]\rho_j\right\rangle+\braket{\left(P_0-\alpha\rho_j+\Gamma\right) \rho_j}\nonumber\\
&{-}2\left\langle\left[ \left(P_0-\alpha\rho_j\right)\left(\rho_j+\frac12\right)-\Gamma \left(\rho_j - \frac12\right)\right]\right\rangle\braket{\rho_j}=2\left[P_0-\Gamma-\alpha\right]C_j+(P_0+\Gamma)\braket{\rho_j}-2\alpha{\rm Cov}(\rho_j, \rho_j^2) -\alpha\braket{\rho_j}^2,\nonumber\\
&d_tC_{12}=\frac{d\braket{\rho_1\rho_2}-d(\braket{\rho_1}\braket{\rho_2})}{dt}=\left\langle\left[ \left(P_0-\alpha\rho_1\right)\left(\rho_1{+}\frac12\right){-}\Gamma \left(\rho_1 {-} \frac12\right)\right]\rho_2{+}\left[ \left(P_0{-}\alpha\rho_2\right)\left(\rho_2{+}\frac12\right)-\Gamma \left(\rho_2 {-} \frac12\right)\right]\rho_1\right\rangle\nonumber\\
&-\left\langle\left[ \left(P_0-\alpha\rho_1\right)\left(\rho_1{+}\frac12\right){-}\Gamma \left(\rho_1 {-} \frac12\right)\right]\right\rangle\braket{\rho_2}{-}\left\langle\left[ \left(P_0{-}\alpha\rho_2\right)\left(\rho_2{+}\frac12\right)-\Gamma \left(\rho_2 {-} \frac12\right)\right]\right\rangle\braket{\rho_1}=\nonumber\\
&=\left[2\left(P_0-\Gamma\right)-\alpha\right]C_{12}-\alpha[{\rm Cov}(\rho_1^2, \rho_2)+{\rm Cov}(\rho_2^2,\rho_1)].
\end{align}
\end{widetext}
As we approximate the joint distribution \eqref{eq:gaussian_potential} of $\rho_1$ and $\rho_2$ as Gaussian, we can use Isserlis' theorem to derive a closed system of equations. Namely:
\begin{align}
    &{\rm Cov}(\rho_j^2, \rho_j)=2\braket{\rho_j}C_j,\\
    &{\rm Cov}(\rho_1^2,\rho_2)+{\rm Cov}(\rho_2^2,\rho_1)=2[\braket{\rho_1}+\braket{\rho_2}]C_{12},
\end{align}
therefore
\begin{align}
d_tC_j&=2[P_0{-}\alpha{-}\Gamma{-}2\alpha\braket{\rho_j}]C_j{+}(P_0{+}\Gamma{-}\alpha\braket{\rho_j})\braket{\rho_j},\nonumber\\
d_tC_{12}&=2\left[\left(P_0-\Gamma\right)-\alpha\left(\braket{\rho_1}+\braket{\rho_2}+\frac12\right)\right]C_{12}.
\end{align}
Within the weak-noise and high-occupation approximations, $\braket{\rho_j}\gg1$ and $\braket{\rho_j}^2\gg C_j, |C_{12}|$, the equations for the mean occupations decouple from the cumulants:
\begin{align}
    d_t\braket{\rho_j}=\left(P_0-\alpha\braket{\rho_j}-\Gamma\right)\braket{\rho_j}
\end{align}
and may be solved explicitly and expressed in terms of the dimensionless time $\tau = \Gamma t$, parameters $\zeta_0=2\chi_0/\Gamma$ and $f=P/P_{\rm th}-1=P_0/\Gamma-1$ from the main text:
\begin{align}\label{eq:appB_rhot}
    \braket{\rho_1}(\tau)=\braket{\rho_2}(\tau)=\rho_m(\tau)=\frac{f\Gamma}{\alpha}\frac{f+\zeta_0}{f+\zeta_0(1-e^{-f\tau})}.
\end{align}
Here we took advantage of the symmetry of the initial distribution \eqref{eq:Wigner_stationary} and used \eqref{eq:equilibrium_rho_cond} to define the initial conditions.

Within the same approximations (high occupation and weak noise limit), the equations for cumulants read as follows ($C(\tau)=C_1(\tau)=C_2(\tau)$ due to the symmetry of the equations and the initial distribution):
\begin{align}
    d_\tau C&=2\left(f-2\frac{\alpha\rho_m}{\Gamma}\right)C+\left(f+2-\frac{\alpha\rho_m}\Gamma\right)\rho_m,\\
 d_\tau C_{12}&=2\left(f-2\frac{\alpha \rho_m}\Gamma\right)C_{12}.
\end{align}
Given $\rho_m(\tau)$, these equations can be integrated explicitly:
\begin{align}
C(\tau)&=h(\tau)\left[C(0)+\int_0^\tau\frac{\rho_m(s)}{h(s)}\left[f+2-\frac{\alpha \rho_m(s)}{\Gamma}\right]ds\right],\label{eq:appb_ct}\\
C_{12}(\tau)&=h(\tau)C_{12}(0)\label{eq:appb_c12t}    
\end{align}
with
\begin{align}
    h(\tau)=e^{\int_0^\tau2\left[f-2\frac{\alpha\rho_m(s)}{\Gamma}\right]ds}=\left[\frac{f}{f +\zeta_0 (1-e^{-f\tau}) }\right]^4e^{-2f\tau}.
\end{align}
The first expression can be expressed as follows:
\begin{align}
C(\tau)&=C(0)h(\tau)+\frac{\Gamma}{\alpha}e(\tau),
\end{align}
with $e(\tau)$ interpolating between $e(0)=0$ and $e(\infty)=1$:
\begin{widetext}
\begin{align}
    e(\tau)=\left[\frac{f+\zeta_0}{f +\zeta_0 (1-e^{-f\tau})}\right]^4\left[\left(1-e^{-f \tau }\right)\left[1-\frac{ f (\zeta_0 -1)+5 \zeta_0
    }{f+\zeta_0}e^{-f\tau}-\frac{(f+2)\zeta_0^3}{(f+\zeta_0)^3}e^{-2f\tau}\right]+\frac{2\zeta_0 ^2 f(f+3)  e^{-2f \tau }\tau}{(f+\zeta_0)^2}\right].
\end{align}
\end{widetext}
Finally, we derive the equation for the angular variance. Since the evolution Eq.~\eqref{eq:app_angular_Ito} describes pure diffusion, it is straightforward. For each of the angles:
\begin{align}
    dC_{\theta_i}=\frac{1}{4}\frac{\Gamma(f+2)-\alpha \rho_m(\tau)}{\Gamma\rho_m(\tau)}d\tau.
\end{align}
Since the noise increments are independent, $d\Omega_1d\Omega_2=0$, $dC_{\theta_+}=dC_{\theta_1}+dC_{\theta_2}$ and
\begin{align}\label{eq:appb_ctht}
    C_{\theta_+}&=C_{\theta_+}(0)+\frac{1}{2}\int_0^\tau\left(\frac{f+2}{\rho_m(z)}-\frac{\alpha}{\Gamma}\right)dz\\
    &=C_{\theta_+}(0)+\frac{2\alpha}{\Gamma}\left[-\frac{(f+2)\zeta_0(1-e^{-f\tau})}{4f^2(f+\zeta_0)}+\frac{\tau}{2f}\right].\nonumber
\end{align}
\section{Estimate of the disentanglement time}\label{app:C}
The initial value of the squeezing parameter for the linear saturation model is given by
\begin{align}
    \xi_0&=1-\frac1{4\rho_m^2}\frac{(2\Gamma{-}2\chi_0)(\alpha \rho_m)}{2\alpha(\alpha\rho_m+2\chi_0)}-\frac{2\Gamma-2\chi_0}{16\chi_0\rho_m}=\nonumber\\
    &=1-\frac1{4}\frac{2{-}\zeta_0}{2\rho_m(\frac{\alpha\rho_m}{\Gamma}+\zeta_0)}-\frac{2-\zeta_0}{8\zeta_0\rho_m}=\nonumber\\
    &=1-\frac{2-\zeta_0}{8\rho_m}\left[\frac{1}{f+2\zeta_0}+\frac{1}{\zeta_0}\right].
\end{align}
Here we substituted $\eta=1$, $\rho_m=(P_0-\Gamma+2\chi_0)/\alpha$ and $G'[\rho_m]=\alpha$. The initial state is entangled if the inequality \eqref{eq:squeezing_inequality} is satisfied, which leads to a threshold value:
\begin{align}
    \zeta^{\rm thr}_0=\frac1{22}\left[6-5f+\sqrt{36+28f+25f^2}\right].
\end{align}
One may check that this is the same result as one would obtain by solving Eqs.~\eqref{eq:critpump_eta1} and \eqref{eq:kappa_linear}.

Analyzing the solutions \eqref{eq:appB_rhot}, \eqref{eq:appb_ct}, \eqref{eq:appb_c12t} and \eqref{eq:appb_ctht}, we see that the radial cross-correlations and transient radial fluctuations decay exponentially, while the angular variance grows only linearly at large times:
\begin{align}
    C_{\theta_+}(\tau \to \infty)=\frac{\alpha}{\Gamma}\left[\underbrace{\frac{2-\zeta_0}{4\zeta_0(f+\zeta_0)}}_{\rm initial\; value}-\frac{(f+2)\zeta_0}{2f^2(f+\zeta_0)}+\frac{\tau}{f}\right].
\end{align}
So, at a late time $\tau\gg1/f$, radial correlations have already faded out, and the squeezing degree is given by the following expression:
\begin{align}
    &\xi(\tau)=1-\frac{C(\infty)}{4\rho_m^2(\infty)}-\frac{C_{\theta_+}(\tau\to \infty)}{2}\\
    &=1-\frac{\alpha}{\Gamma}\left[\frac{1}{4f^2}+\frac{f^2(2-\zeta_0)-2\zeta_0^2(f+2)}{8f^2\zeta_0(f+\zeta_0)}+\frac{\tau}{2f}\right].\nonumber
\end{align}
Once the radial distribution has relaxed towards the new equilibrium, the entanglement threshold is given by
\begin{align}   
\xi_{\rm cr}=1-\frac1{2\rho_m(\infty)}=1-\frac{\alpha}{\Gamma}\frac{1}{2f},
\end{align}
thus, the entanglement criterion is as follows:
\begin{align}
    \left[\frac{1}{4f^2}+\frac{f^2(2-\zeta_0)-2\zeta_0^2(f+2)}{8f^2\zeta_0(f+\zeta_0)}+\frac{\tau}{2f}\right]&<\frac{1}{2f}.
\end{align}
Within the late-time approximation, it is satisfied while $\tau<\tau_d$ with
\begin{align}
\tau_d=\frac{1}{2}\left(\frac52-\frac{1}{\zeta_0}\right)+\frac{\zeta_0}{4f}\frac{f+2}{f+\zeta_0}.
\end{align}
Note that $\zeta_0=2/5$ is the minimum possible value of the entanglement boundary, see Eq. \eqref{eq:critpump_eta1}, which ensures the positivity of this expression. Numerical evaluation using the exact solutions \eqref{eq:appB_rhot}, \eqref{eq:appb_ct}, \eqref{eq:appb_c12t}, and \eqref{eq:appb_ctht} shows that the above late-time expression provides an upper bound for the exact critical time. For fixed $\zeta_0>2/5$, the leading-order large-$f$ asymptotics of the exact disentanglement time is
\begin{align}
\tau_d^{\rm exact}=\frac12\left(\frac{5}2-\frac1{\zeta_0}\right)+O(f^{-1}).
\end{align}
This asymptotics is valid provided that $f(\zeta_0-2/5)\gg1$, ensuring that radial relaxation is completed before disentanglement.
\end{document}